\def   \ni {\noindent}

\def   \ssk {\vskip  5truept}

\def   \bsk {\vskip 15truept}
 
\def   \newpage {\vfill\eject}
\def   \newline {\hfil\break}

\def\beq{\begin{equation}}
\def\eeq{\end{equation}}
\newcommand*{\AB}{{\mathbf A}}
\newcommand*{\MB}{{\mathbf M}}
\newcommand*{\NB}{{\mathbf N}}
\newcommand*{\OB}{{\mathbf O}}
\newcommand*{\DB}{{\mathbf \tilde{O}}}
\newcommand*{\RB}{{\mathbf R}}
\newcommand*{\SB}{{\mathbf S}}
\newcommand*{\SC}{{\mathbf \tilde{S}}}
\newcommand*{\XB}{{\mathbf X}}
\newcommand*{\da}{\delta}
\newcommand*{\RLM}{\RB\left[\psi(l,m,\da)\right]}
\newcommand*{\id}{{\mathbf 1}}

\documentstyle[epsfig]{article}
\begin{document}

%
\def\la{\mathrel{\mathchoice {\vcenter{\offinterlineskip\halign{\hfil
$\displaystyle##$\hfil\cr<\cr\sim\cr}}}
{\vcenter{\offinterlineskip\halign{\hfil$\textstyle##$\hfil\cr
<\cr\sim\cr}}}
{\vcenter{\offinterlineskip\halign{\hfil$\scriptstyle##$\hfil\cr
<\cr\sim\cr}}}
{\vcenter{\offinterlineskip\halign{\hfil$\scriptscriptstyle##$\hfil\cr
<\cr\sim\cr}}}}}
\def\ga{\mathrel{\mathchoice {\vcenter{\offinterlineskip\halign{\hfil
$\displaystyle##$\hfil\cr>\cr\sim\cr}}}
{\vcenter{\offinterlineskip\halign{\hfil$\textstyle##$\hfil\cr
>\cr\sim\cr}}}
{\vcenter{\offinterlineskip\halign{\hfil$\scriptstyle##$\hfil\cr
>\cr\sim\cr}}}
{\vcenter{\offinterlineskip\halign{\hfil$\scriptscriptstyle##$\hfil\cr
>\cr\sim\cr}}}}}
\def\degr{\hbox{$^\circ$}}
\def\arcmin{\hbox{$^\prime$}}
\def\arcsec{\hbox{$^{\prime\prime}$}}

\hsize 5truein
\vsize 8truein
\font\abstract=cmr8
\font\keywords=cmr8
\font\caption=cmr8
\font\references=cmr8
\font\text=cmr10
\font\affiliation=cmssi10
\font\author=cmss10
\font\mc=cmss8
\font\title=cmssbx10 scaled\magstep2
\font\alcit=cmti7 scaled\magstephalf
\font\alcin=cmr6 
\font\ita=cmti8
\font\mma=cmr8
\def\ref{\par\noindent\hangindent 15pt}
\null


\title{\ni Destriping Polarised data
}                                               

\bsk \bsk
\author{\ni B. Revenu $^1$, F. Couchot $^2$, J. Delabrouille
  $^{3,\,1}$, \underline{J. Kaplan}$^1$}                                                       
\bsk
\affiliation{1)  Physique Corpusculaire et Cosmologie, Coll{\`e}ge de
  France, 11 Place Marcelin Berthelot, 75231 Paris Cedex 05, France
}

\affiliation{2) Laboratoire de l'Acc{\'e}l{\'e}rateur Lin{\'e}aire, IN2P3 CNRS, Universit{\'e} 
Paris Sud, 91405 Orsay, France
}

\affiliation{3) Institut d'Astrophysique Spatiale, CNRS \& Universit{\'e}
  Paris XI, b{\^a}t 121, 91405 Orsay Cedex, France.
}                                                
\bsk
\baselineskip = 12pt

\abstract{ABSTRACT \ni
We show how to take advantage of a circle scanning strategy, such as
that planned for the Planck mission, to get rid of low frequency
noises for polarised data.
}                                                    
\bsk
\baselineskip = 12pt
\keywords{\ni KEYWORDS: 
}               

\bsk
\baselineskip = 12pt


\text{\ni 1. INTRODUCTION
\ssk
\ni     

Polarisation measurements on the Cosmological Microwave Background (CMB) will
be very difficult as the expected polarisation is not expected to
exceed 10\% of the signal. Therefore, noises of all origins must be
searched for and carefully extracted. 

Low frequency drifts are expected to arise from mechanic, 
thermal and electronic origin. If not suitably treated they induce
stripes on sky maps such as those shown on figure 1.
\begin{figure}[h]
\centerline{\epsfig{file=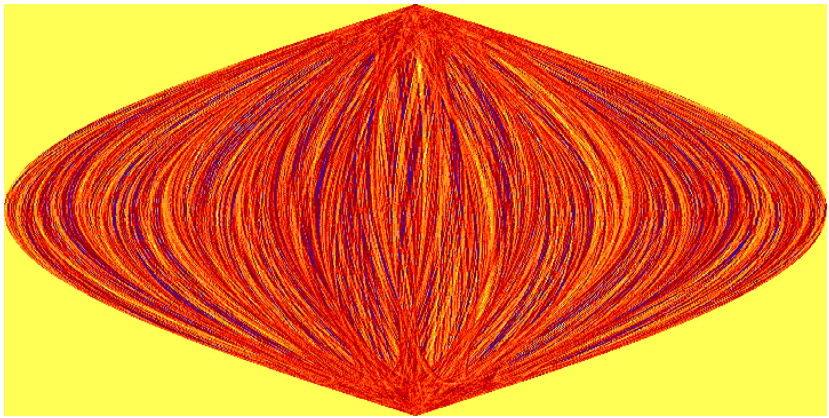, width=.9\textwidth}}
\centerline{\caption{FIGURE 1. Stripes on the sky map of the $Q$ Stokes
  parameter (no signal).}
}
\end{figure}
The elimination of low frequency noise has been already studied in
the framework of differential measurements as was done in COBE DMR and 
as planed for the MAP mission (Janssen and Gulkis, 1992; Wright et
al. 1996; Tegmark 1997). A solution to this problem in the case of
circle scanning strategies has been proposed by
Delabrouille (1998a) for temperature measurements.  We
discuss here how to extend this procedure to polarised data. 
\bsk
\ni 2. PRINCIPLE OF THE METHOD
\ssk
\ni 
The destriping method uses the two kind of redundancies provided by a
circle scanning strategy:\\
1) The
  telescope scans some 60 times a fixed circle in the sky with an
  angular radius of 
  $\sim 85^\circ$, at a frequency   $f_{\mathrm{spin}} \sim 1\
  \mathrm{rpm}$. The signal along one circle can be obtained in a
  nearly optimal way by averaging 
  all 60 scans. As a result, all non synchronous 
  noises with frequencies smaller than $f_{\mathrm{spin}}$ reduce to
  one offset per circle in a first approximation. These offsets
  are the main source of the stripes in figure 1.\\
2) Every hour, the spin axis of the mirror is moved by a few
arcminutes. The different scanning circles have many intersections
in the sky. At any point where two circles cross each other, the
  physical signal must be the same along both of them, and the 
  resulting contraints allow to determine and subtract the offsets. 
However at intersection points, a given polarimeter has
different directions along the two circles, as shown in figure 2,  and
therefore records different signals if the polarisation is non zero.
\begin{figure}[h]
\centerline{\epsfig{file=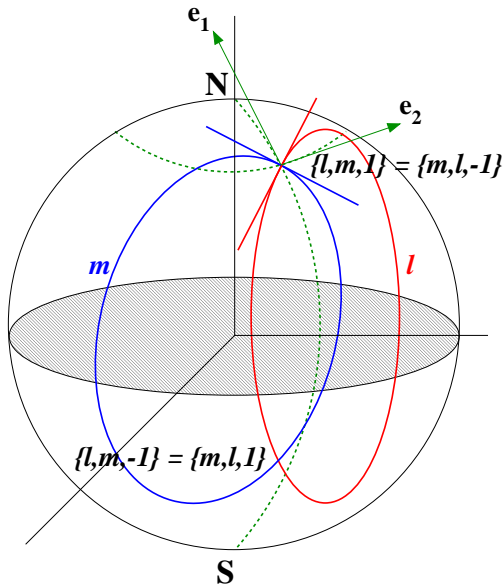,  width=.55\textwidth}}
\caption{FIGURE 2 The intersections of two scanning circle. Here, the
  polarimeter is assumed to be tangent to the scanning circle. 
}
\end{figure}
The solution is to compare the Stokes parameters $I$, $Q$ and $U$
measured along the two circle, but defined in a fixed global reference frame,
for instance the ecliptic latitude-longitude frame (${\mathbf e_1},
{\mathbf e_2}$) drawn in figure 2.
\bsk
\ni 3. APPLICATION TO POLARISED DATA
\ssk
\bsk
\ni
3.1 The relation between polarimeter outputs and Stokes parameters
\ssk
\ni
\begin{figure}[h]
\centerline{\epsfig{file=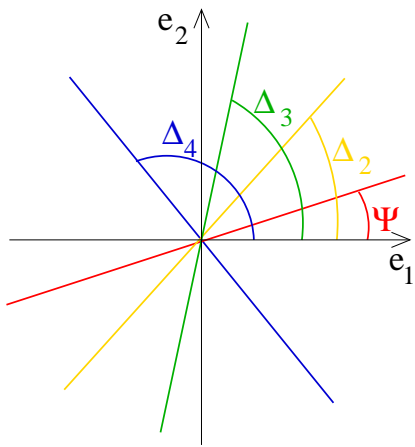,angle=0, width=.45\textwidth}}
\caption{FIGURE 3. The setup of the polarimeters in the focal plane
  and the angle of rotation $\psi$ with respect to the fixed reference 
  frame (${\mathbf e_1}, {\mathbf e_2}$)
}
\end{figure}
If the focal plane involves $n$ polarimeters in a given wavelength, their outputs can be described by a vector $\MB$
with $n$ components, and the Stokes parameters in the fixed global
reference frame by another vector $\SB$ with
3 components:
\[
 \MB = 
\left(\begin{array}{c}
m_1\\ \vdots \\m_p\\ \vdots \\m_n 
\end{array}\right), \mbox{ and }
\SB=\left(\begin{array}{c} I\\Q\\U \end{array}\right)\mbox{, related by
  } \MB = \AB\ \RB(\psi)\ \SB
\]
The matrix $\AB$ characterises the setup of the polarimeters in 
  the focal plane, whereas the rotation $\RB(\psi)$ depends on the
  orientation of the focal plane with respect to the the fixed global
  reference frame: 
\[
  \AB =\frac{1}{2} \left(\begin{array}{ccc}
    1&1&0\\
    \vdots & \vdots & \vdots\\
    1&\cos 2 \Delta_p&\sin 2 \Delta_p\\
    \vdots & \vdots & \vdots\\
    1&\cos 2 \Delta_{n-1}&\sin 2 \Delta_{n-1}\\
  \end{array}\right)
\mbox{, and }
  \RB(\psi)=\left(\begin{array}{ccc}
      1&0&0\\
      0&\cos 2\psi&\sin 2\psi\\
      0&-\sin 2\psi&\cos 2\psi 
    \end{array}\right). \nonumber
\]
The meaning of angles $\Delta_p$ and $\psi$ is shown in figure 3
\bsk
\ni
3.2 Determination of the offsets
\ssk
\ni
As shown in figure 2,
each scanning circle $l$ has two intersections with circle $m$, denoted 
by  $\{l,m,\da=\pm 1\}$. Intersections $\{l,m,\da\}$ and $\{m,l,-\da\}$
correspond to the same point in the sky.
To determine the offsets, one uses the fact that the Stokes parameters 
are uniquely defined in the fixed global reference frame:
\beq
\SB_{l,m,\da} = \SB_{m,l,-\da} 
\eeq
From the vector of polarimeter outputs $\MB$, one gets the Stokes parameters at the
intersections and the offsets by minimizing a $\chi^2$, taking
relation (1) into account. After eliminating the Stokes parameters,
the minimisation conditions reduce to the following equation: 
\begin{equation}
\begin{array}{c}
\sum_{m,\da}\left[\id + \RLM\,
  {\XB_m}^{-1}\,\RLM^{-1}\,\XB_l\right]^{-1} \times ~~~~~~~~~~~~~~~~~~~~~~~~ \\
 ~~~~~~~~~~~~~\left[\DB_l - \RLM\,\DB_m - 
\SC_{l,m,\da} + \RLM\,\SC_{m,l,-\da}\right]  = 0.
\end{array}
\end{equation} 
In Equation (2) 
$\SC_{l,m,\da} = \XB_l^{-1}\,\AB^T\, {\NB_l}^{-1} \MB_{l,m,\da}$ 
is known in terms of the vectors of polarimeters
outputs $\MB_l$ along circle $l$  and their noise matrix $\NB_l$
(assumed to be white after the averaging procedure).  $\SC_{l,m,\da}$
can be interpreted as the vector of the Stokes parameters in the focal
reference frame, and $\XB_{l} = \AB^T {\NB_{l}}^{-1} \AB$ is the
inverse of the associated noise matrix. Also known is the rotation
matrix $\RLM$ wich relates the focal frame Stokes parameters along circle
$m$ to those along circle $l$.

The unknown offsets appear through the combination 
$\DB_l = {\XB_l}^{-1}\,\AB^T\, {\NB_l}^{-1} \OB_l$, which can be
viewed as the offsets on the focal frame Stokes parameters. 

Once the focal frame offsets $\DB_l$ are determined by solving equation (2),
one can reconstruct the fixed global frame Stokes parameters along all
circles as:
\[
\SB_{l,k} = \RB_{l,k} (\SC_{l,k} - {\DB_l}),
\]
where rotation $\RB_{l,k}$ transforms the Stokes
  Parameters from the focal to the fixed global reference frame at pixel $k$ 
  along circle $l$.
\newpage
\bsk
\ni 4. DISCUSSION
\ssk
\ni
Figure 4 shows the same map as figure 1, after application of the
\begin{figure}[h]
\centerline{\epsfig{file=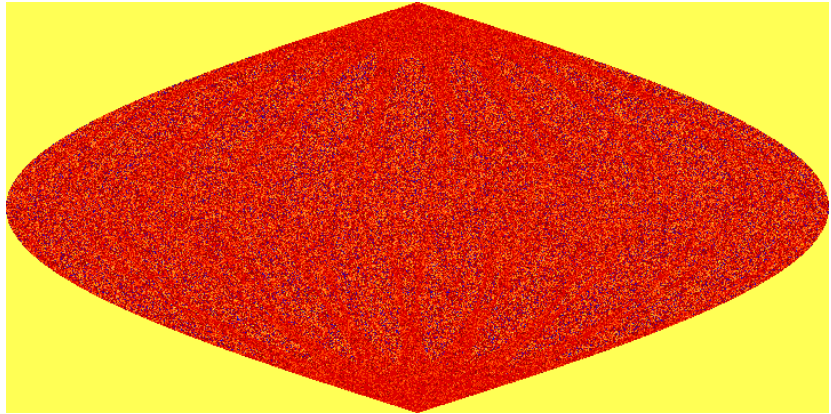, width=.9\textwidth}}
\centerline{\caption{FIGURE 4. The sky map of the $Q$ Stokes
  parameter after destriping (no signal).}
}
\end{figure}
destriping procedure described above. The stripes have totally
disapeared. The barely visible radial structures around the poles are
not stripes but  
regions where the level of noise is smaller because the density of
intersection points is larger. Figure 5 shows the density of
intersection points, which corresponds to
\begin{figure}[h]
\centerline{\epsfig{file=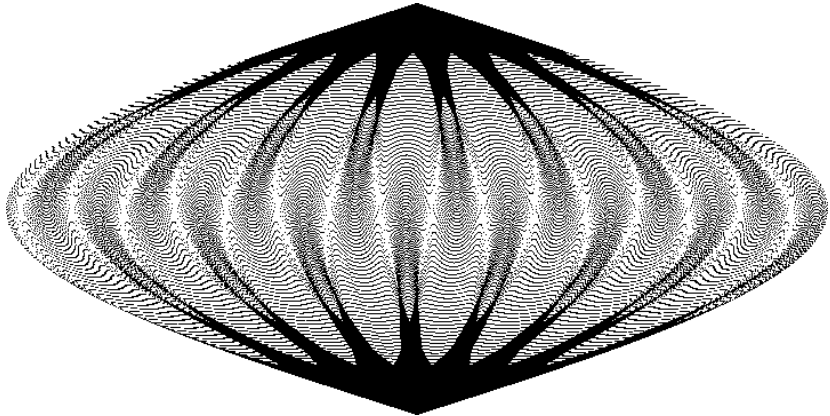, width=.9\textwidth}}
\centerline{\caption{FIGURE 5. The density of intersection points. Dark regions
  correspond to high densities}
}
\end{figure}
a scanning strategy where the spin axis moves along a
sinusoid around the ecliptic plane.

The algorithm developed above to destripe polarised data is already 
as efficient as for unpolarised data, even if several points have
still to be settled, of which we name a few:\\ 
1) Synchronous noises have to be eliminated and this has been treated
at this meeting by Delabrouille et al. (1998b).\\
2) The quality of the destriping has to be tested in a more
sophisticated way, and we are currently developping methods to do so.\\
3) Measurements along two different circles will in reality be different
for other reasons, such has the time response of the electronics and
the asymmetry of the field of view. We are studying ways to modelise
this pheneomenon. \\ 
4) Sidelobes might cause difficulties in polarisation measurements
because light rays with a large incidence angle could carry a large
instrumental polarisation.

}



\bsk
\baselineskip = 12pt


{\references \ni REFERENCES
\ssk

\ref Delabrouille, J. 1998a, A \& A Supplement Series 127, 555
\ref Delabrouille, J. et al. 1998b, these proceedings. 
\ref Janssen, M. A. and Gulkis, S. 1992, The Infrared and
  Submillimeter Sky after {\it COBE}, in Proceedings of the NATO Advanced Study
  Institute, M. Signore and C. Dupraz (eds.), Dordrecht: Kluwer, p. 391,
\ref Tegmark, M. 1997, ApJ Letters 480, L87
\ref Wright, E.~L., Hinshaw, G., and Bennett, C.~L. 1996, ApJ 458, L53

}                      

\end{document}